\documentclass[%
  reprint,
  color,
 amsmath,amssymb,
 aps,
 prd,
 twocolumn,
 float,
 longbibliography,
 nofootinbib
]{revtex4-2}

\usepackage{graphicx}
\usepackage{dcolumn}
\usepackage{bm}
\usepackage{empheq}

\newcommand{\omio}{\Omega_{\rm i}}
\newcommand{\omel}{\Omega_{\rm e}}
\newcommand{\omo}{\Omega_{\rm n}}

\newcommand{\mpro}{m_{\rm p}}
\newcommand{\mio}{m_{\rm i}}
\newcommand{\mel}{m_{\rm e}}
\newcommand{\mo}{m_{\rm n}}

\newcommand{\npro}{n_{\rm H}}
\newcommand{\nio}{n_{\rm i}}
\newcommand{\nel}{n_{\rm e}}
\newcommand{\no}{n_ {\rm n}}

\newcommand{\re}{a}
\newcommand{\rhoc}{\rho_{\rm c}}

\newcommand{\const}{{\rm const.}}
\newcommand{\psiint}{\Psi_{\rm int.}}

\newcommand{\omb}{\Omega_{\rm B}}
\newcommand{\ombo}{\omega_{\rm B}}
\newcommand{\omeff}{\Omega_{\rm eff}}

\newcommand{\bself}{B_{\rm self.}}

\newcommand{\bare}{\bar{\varepsilon}}
\newcommand{\pamb}{p_{\rm amb}}

\newcommand{\xe}{x_{\rm e}}
\newcommand{\mnras}{Mon. Not. of the Royal Astr. Society}
\newcommand{\aap}{Astron. \& Astrophys.}
\newcommand{\apss}{Astrophys. \& Space Science}
\newcommand{\apjl}{Astrophys. J. Letters}
\newcommand{\apjs}{Astrophys. J. Suppl. Series}
\newcommand{\araa}{Ann. Rev. of Astron. \& Astrophys.}

\begin{document}

\preprint{APS/123-QED}

\title{
  The Maclaurin spheroid in disguise.\\New figures of equilibrium with external magnetic support}

\author{Jean-Marc Hur{\'e}}
\author{Cl\'ement Staelen}%
\affiliation{Laboratoire d'Astrophysique de Bordeaux, Univ. Bordeaux, CNRS,\\ B18N, all\'ee Geoffroy Saint-Hilaire, 33615 Pessac, France.}

\date{\today}


\begin{abstract}
We show that a rigidly rotating, homogeneous ellipsoid of revolution threaded by a uniform, coaxial magnetic field is a possible figure of equilibrium. While the spheroidal shape is fully preserved, the rotation rate is modified. Accordingly, we extend the fundamental formula by Maclaurin. In contrast with the non-magnetic case, prolate shapes are permitted, but there are critical states in the form of maximum elongations, depending on ionisation fraction, ion/electron drift, magnetic field and mass-density. As checked from numerical simulations based on the Self-Consistent-Field method, prolate states survive to gas compressibility. The relevance to interstellar clouds is outlined.
\end{abstract}


\maketitle

\section{The Maclaurin spheroid}
\begin{figure}
       \centering
       \includegraphics[width=8.7cm,trim={0.4cm 0.cm 0.cm 0.cm},clip]{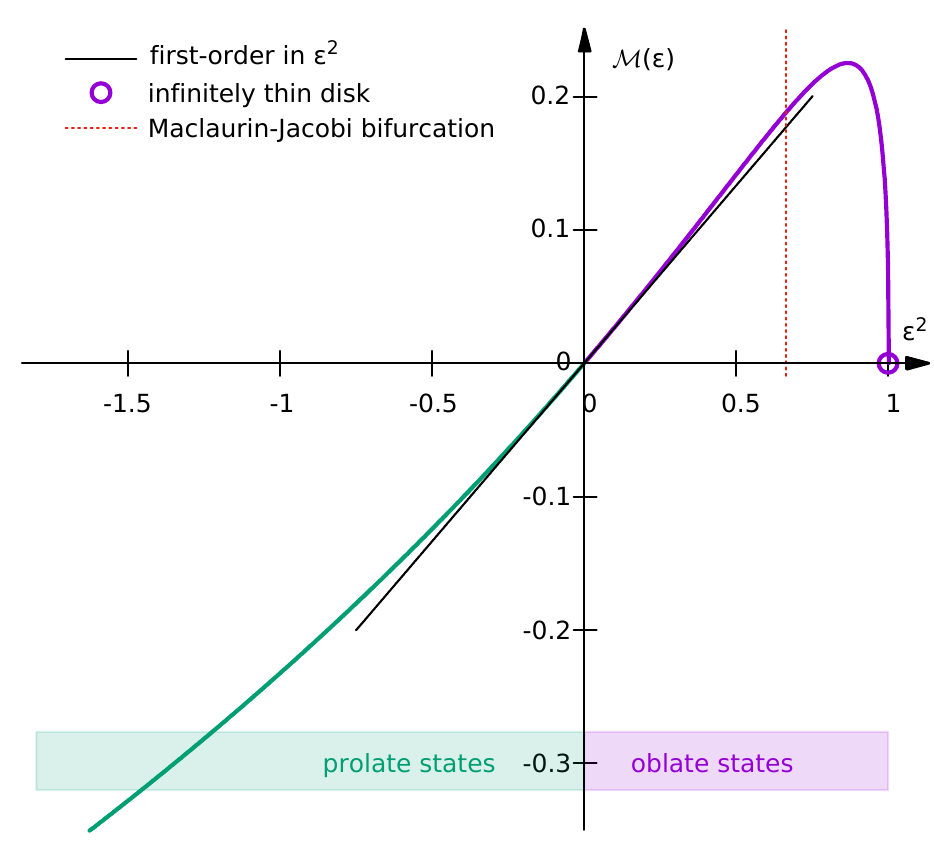}
       \caption{The function ${\cal M}$ defined by Eq.\eqref{eq:mlfunction} versus $\varepsilon^2$ and its first-order approximation in $\varepsilon^2$.}
       \label{fig:maclaurin.pdf}
\end{figure}

Inside a homogeneous ellipsoid of revolution (i.e. a spheroid) rigidly rotating in the vaccum around its minor axis, the gravitational force is exceptionally linear in the cylindrical radius $R$, which exactly compensates the centrifugal force \citep[][]{chandra69,binneytremaine87}. The relationship between the rotation rate $\Omega$, the mass density $\rho$ and the eccentricity $\varepsilon$ of the body has been established by Maclaurin \cite{maclaurin42}
\begin{subequations}
    \begin{empheq}[left={\empheqlbrace}]{align}
  &\frac{\Omega^2}{2\pi G \rho} = {\cal M}(\varepsilon), \label{eq:maclaurin}\\
  &{\cal M}(\varepsilon)=\left(3-2\varepsilon^2\right)\frac{\bare}{\varepsilon^3}\arcsin \varepsilon+3-\frac{3}{\varepsilon^2}, \label{eq:mlfunction}
     \end{empheq}
\end{subequations}
where $a$ and $b$ are the semi-major and semi-minor axis of the surface ellipse respectively, $\varepsilon^2=1-\bare^2$ is the square of the eccentricity, $\bare=b/a$ and $G$ is the gravitation constant. This is a fundamental result for stellar, planetary and even galactic Astrophysics. According to the classical theory, only oblate shapes defined by $\bare\le1$ are possible, which is corroborated by most observations (self-gravitating systems are flattened by rotation). For fluids, the pressure $p$ is found from the Bernoulli invariant \citep{lyttleton1953stability}
\begin{flalign}
  \frac{p}{\rho} - \frac{1}{2}\Omega^2 R^2 + \psiint = \const,
  \label{eq:bernoulli}
\end{flalign}
where the constant can be evaluated at any point $(R,Z)$ in the spheroid, and $\psiint$ is the interior gravitational potential. It takes the form
\begin{flalign}
\psiint(R,Z)= -\pi G \rho\left(A_0a^2-A_1R^2 -A_3Z^2\right)
  \label{eq:psiint},
\end{flalign}
where the coefficients $A_i$'s are known fonctions of $\varepsilon$ \citep[][]{chandra69,binneytremaine87}. For prolate shapes, $\bare$ exceeds unity and $\varepsilon$ becomes a pure imaginary number, and Eq.\eqref{eq:psiint} still holds. The reason resides in the mathematical continuity of the $A_i$'s at $\varepsilon=0$ when crossing over the spherical shape. Actually,  for any complex number $z$, we have $\arcsin(z) = {\rm arcsinh}({\rm i}z)/{\rm i}$, where $\rm i$ is the imaginary unit \citep[][]{gradshteyn15}. As a matter of fact, ${\cal M}(\varepsilon)$ remains a real number as can be seen in Fig. \ref{fig:maclaurin.pdf}, but the function is negative. According to \eqref{eq:maclaurin}, this leads to unphysical states since neither $\Omega^2$ nor $\rho$ can be negative.

\section{A homogeneous spheroid threaded by a coaxial, uniform magnetic field}
\label{sec:magn}

If the homogeneous spheroid is threaded by a magnetic field $\bm{B}$, free charges experience the Lorentz force. The modification of the dynamics is expected to be weak if $B$ is small enough or/and if the amount of charges is relatively small with respect to neutral species. In order to the study this effect, we assume for simplicity that the medium contains electrons and atoms of a single kind capable of releasing $Z \ge 1$ electrons each, and all moving around the minor axis of the system. We denote $n_\alpha$, $m_\alpha$ and $\bm{V}_\alpha=\Omega_\alpha R {\bm e}_\phi$ the number density, mass and circular velocities of particles respectively, with $\alpha \in \{\text{e},\text{i},\text{n}\}$ for electrons, ions and neutrals. The local electro-neutrality implies $\nel=Z \nio$, and so the net current density is
\begin{equation}
  \bm{j}=\nel (-e)  {\bm V}_{\rm e}+ \nio (+Ze) {\bm V}_{\rm i}  = -e \nel (\omel-\omio)R \bm{e}_\phi,
  \label{eq:jcurrent}
  \end{equation}
where $e>0$ is the elementary charge. A single-fluid description is complicated and requires some prescriptions or extra-assumptions, in particular regarding the velocity drift appearing in Eq.\eqref{eq:jcurrent} and the link with the kinematics of neutrals which are pushed through collisions. By adding the individual Euler equations together, we can get rid of momentum exchanges between particles \citep[e.g.][]{benz2002}. For steady states, we have
\begin{equation}
\sum_\alpha \left[n_\alpha m_\alpha (-R\Omega^2_\alpha \bm{e}_\phi + \bm{\nabla} \psiint) + \bm{\nabla} p_\alpha \right] - \bm{L}=\bm{0},
  \label{eq:eulers}
  \end{equation}
where $\bm{L} \equiv \bm{j} \times \bm{B}$ is the total Lorentz force. As Eq.\eqref{eq:eulers} is intended to be recast in the form of Eq.\eqref{eq:bernoulli}, we need a definition for the ``typical'' circular velocity $V$ or for the bulk rotation rate $\Omega$. We could use the standard barycentric formulation $\rho {\bm V} \equiv \sum_\alpha{n_\alpha m_\alpha {\bm V}_\alpha}$, but it seems preferable to work with the centrifugal potential, which is represented by the second term in the left-hand-side of Eq.\eqref{eq:bernoulli}. The main reason is to ensure some continuity with the (non-magnetic) Maclaurin case in the limit where $B \rightarrow 0$. As $Z$ cannot take large values in a standard astrophysical context ($Z=1$ for a pure hydrogen gas) and given the ratio $\mel/\mo \ll 1$, we can omit the contribution of electrons in the non-inertial term in Eq.\eqref{eq:eulers}. We then define $\rho \Omega^2 \equiv \sum_\alpha{n_\alpha m_\alpha \Omega_\alpha^2}$, which leads to
\begin{equation}
  \Omega^2 \equiv \frac{\xe}{\xe+Z} \omio^2+ \frac{Z}{\xe+Z} \omo^2,
  \end{equation}
where $\xe=\nel/\no$ is the ionization fraction, and
\begin{equation}
  \rho = \nel \mio \frac{Z+\xe}{Z \xe}.
  \label{eq:rhomean}
\end{equation}

It follows that Eq.\eqref{eq:eulers} becomes
    \begin{equation}
-R\Omega^2 \bm{e}_\phi + \bm{\nabla} \psiint + \frac{1}{\rho}\bm{\nabla} p -  \frac{\bm{L}}{\rho}=\bm{0}.
  \label{eq:eulers2}
    \end{equation}

Regarding the magnetic field, we assume that it is uniform and aligned (parallel or anti-parallel) with the rotation vector, namely $\bm{B}=B \bm{e}_z$, and  remains unperturbed by the rotating charges present in the spheroid. From Eqs.\eqref{eq:jcurrent} and \eqref{eq:rhomean}, the Lorentz force per unit mass is,
\begin{equation}
  \frac{\bm{L}}{\rho} = -\ombo \frac{Z \xe}{Z+\xe}(\omel-\omio)R \bm{e}_R,
\end{equation}
where $\ombo=eB/\mio$ (the sign of this quantity depends on the orientation of the magnetic field). The difference $\omel-\omio$ is hard to guess or anticipate without going into the details of theorical considerations or numerical simulations \citep[e.g.][]{mokc22}. In the present case, it is prescribed as a fraction of the rotation rate, namely
    \begin{equation}
      \omel-\omio \equiv \delta \Omega,
      \label{eq:assumptiondrift}
    \end{equation}
    where the electron/ion drift-parameter $\delta \gtrless 0$ is assumed to be constant in the system. This quantity is probably, statistically, smaller that thermal speeds by orders of magnitude \cite{gfp09}.
      If the Lorentz force does not depend on the $z$-coordinate, then it is the gradient of a magnetic potential $\Lambda$, namely $\frac{1}{\rho}\bm{L} = - \nabla \Lambda$. This is realized, for instance, if $\xe$ and $\delta$ are contant in the body and if the $\Omega_\alpha$'s are constant on cylinders \citep{tassoul78,amendt1989}. For rigid rotations, we basically have
\begin{flalign}
  \Lambda = \frac{1}{2}  \delta \omb \Omega R^2 + \const,
  \label{eq:lambda}
\end{flalign}
where we have set
\begin{equation}
  \omb=\frac{Z\xe}{Z+\xe} \ombo,
\end{equation}
 for convenience. It follows that Eq.\eqref{eq:bernoulli} takes the usual conservative form \citep{te05,lj09,f15,tksk16,dmp21}
\begin{flalign}
  \frac{p}{\rho} - \frac{1}{2}\Omega^2 R^2 + \psiint + \Lambda= \const
  \label{eq:bernoulli_bfield}
\end{flalign}

The fact that $\Lambda$ is quadratic in $R$, like the centrifugal potential, implies that Eq.\eqref{eq:bernoulli} is formally preserved in the magnetic case if the rotation rate $\Omega$ of the Maclaurin spheroid is replaced by an ``effective'' rate $\omeff$, with
\begin{flalign}
  \omeff^2 = \Omega^2\left( 1 - \eta\right),
  \label{eq:om2eff}
\end{flalign}
where
\begin{flalign}
  \eta=\delta\frac{\omb}{\Omega}\gtrless 0.
\end{flalign}

It also means that the spheroidal figure of the equilibrium is fully conserved. This property has been quoted already \citep{ktye11}, but the authors have assumed that the uniform ${\bm B}$-field is created by the star itself. Under the hypothesis made here, we can state that {\it a rigidly rotating, homogeneous spheroid and threaded by a coaxial, uniform magnetic field is a possible figure of equilibrium}.

\section{The new rotation rate}

\begin{figure}
       \centering
       \includegraphics[width=8.7cm,trim={0.4cm 0.cm 0.cm 0.cm},clip]{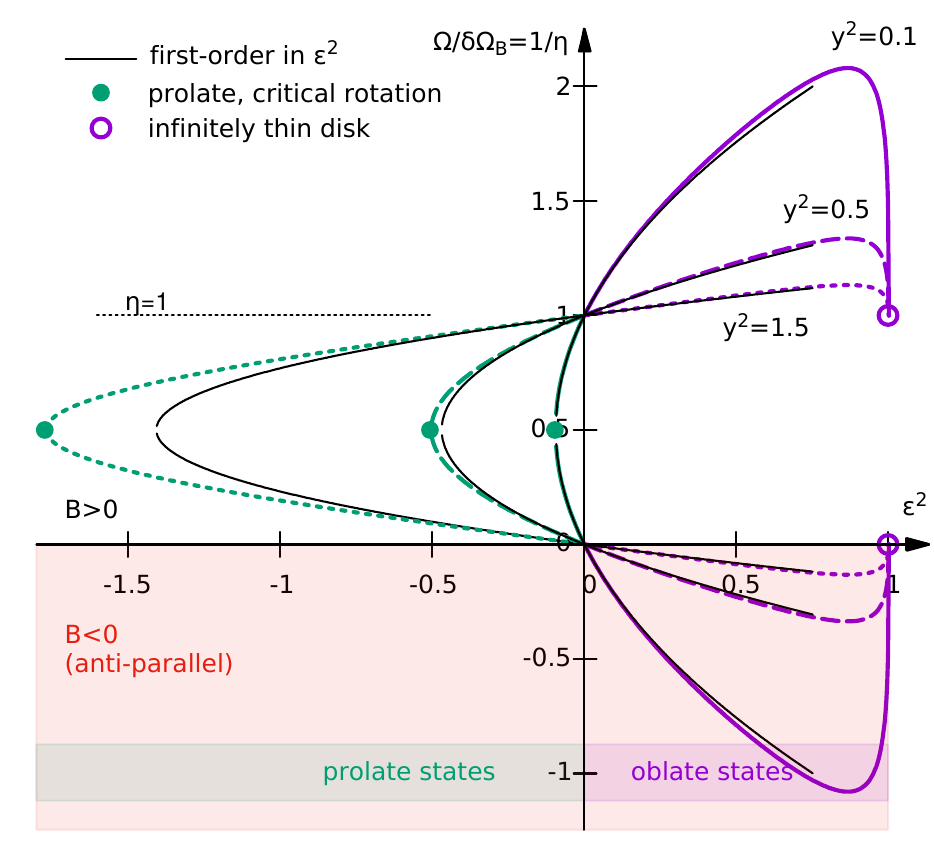}
       \caption{The roots of Eq.\eqref{eq:etaequation} versus the square of the eccentricity of the spheroid $\varepsilon^2$ for $y^2 \in \{0.1, 0.5,1.5\}$.}
       \label{fig:roots.pdf}
\end{figure}

A direct consequence of Eq.\eqref{eq:bernoulli_bfield} is that we can define an ``effective'' function ${\cal M}_{\rm eff}$ from Eq.\eqref{eq:maclaurin} by replacing $\Omega$ by $\omeff$ as given by Eq.\eqref{eq:om2eff}. We see that $\omeff^2$ can be positive or negative depending on the strength of the field, its orientation and on the sign of the electron/ion drift parameter. It means that both oblate and prolate configurations are allowed (see Fig. \ref{fig:maclaurin.pdf}), and $\eta$ plays a critical role (magnitude and sign). When $|\eta| \rightarrow 0$, the impact of the magnetic field is negligeable and one recovers Maclaurin's solution. For given $\delta \omb$, $\rho$ and $\varepsilon$, the rotation rate is calculated from Eq.\eqref{eq:om2eff}
    \begin{equation}
      \Omega^2 - \delta \omb \Omega  -  2\pi G \rho {\cal M}(\varepsilon) = 0,
    \end{equation}
    which is easily expressed in terms of $1/\eta$ by dividing by $\delta^2 \omb^2$. By defining $y= \frac{\delta\omb}{\sqrt{2\pi G \rho}}$, we find
    \begin{equation}
      \frac{1}{\eta^2} - \frac{1}{\eta} - \frac{{\cal M}(\varepsilon)}{y^2} = 0.
      \label{eq:etaequation}
    \end{equation}
This $2$nd-degree equation is easily solved (see the last section for the solution). The roots are plotted versus $\varepsilon^2$ in Fig. \ref{fig:roots.pdf} for three values of the parameter $y^2$.\\

 {\bf Spherical configurations}.| As Eq.\eqref{eq:om2eff} shows, the equilibrium configuration is a sphere in two cases. First, this occurs for $\eta=1$ meaning that the body rotates. This exceptional tuning requires $\delta B>0$, i.e. ${\bf B}$ is aligned with ${\bf \Omega}$ for a positive drift-parameter $\delta$. If $|\delta| \ll 1$, then $\Omega \ll \omb$. The second case is more standard, and corresponds to the absence of rotation $\Omega \rightarrow 0$ (relative to $\delta \omb$), although $|\eta|$ can be very large. \\

 {\bf Oblate configurations}.| Oblate shapes occur with two different orientations of $\bm{B}$, parallel or anti-parallel to $\bm{\Omega}$, as long as $\eta < 1$. There is a solution for any real eccentricity in the range $[0,1]$. In the limit where $\varepsilon \rightarrow 1$, the body tends to an infinitely flat, finite-radius disk, with either $\Omega \rightarrow 0$ rate and finite $\omb>0$, or $\Omega=\delta \omb$.\\

 {\bf Prolate configurations}.| For prolate states, the term inside the parenthesis in Eq.\eqref {eq:om2eff} must be negative, i.e. $\eta >  1$. This situation is new. It is met when the ionization fraction or $\delta B>0$ is high enough, or both. The equilibria are located in the left-side of the graph in Fig. \ref{fig:roots.pdf}. In contrast with the oblate case, not all values of $|\varepsilon^2|$ are allowed in the range $[0,\infty[$. The fact that $\eta$ in Eq.\eqref{eq:etaequation} is necessarily a real number sets a limit in terms of i$\varepsilon$ or, equivalently in terms of the axis ratio $b/a\equiv \bare$. The limit is reached when $\eta$ is a double root (namely, $\eta=2$), which corresponds to
\begin{flalign}
      4 {\cal M}(\varepsilon) = -\underbrace{\delta^2 \omb^2/2\pi G \rho}_{y^2} \le 0,\label{eq:crot}
    \end{flalign}
Physically, this is a {\it critical rotation}. The value of $\epsilon^2$ in Eq.\eqref{eq:crot} is denoted $\varepsilon^2_{\rm cr}$. In other words, there is no magnetic state for spheroids with $\frac{b}{a} > \bare_{\rm cr}$. For slow rotations where $\varepsilon^2 \rightarrow 0^-$, we have ${\cal M}(\varepsilon) \rightarrow \frac{4}{15}\varepsilon^2$, and the critical axis-ratio is given by
    \begin{flalign}
      \bare_{\rm cr}^2 \approx 1+\frac{15}{16}y^2 \ge 1.
    \end{flalign}
    When $\varepsilon^2 \rightarrow -\infty$, the body becomes a filament, and we have ${\cal M}(\varepsilon) \sim 3- \ln(-4\varepsilon^2)$.

    \section{Beyond the incompressible case}

\begin{figure}
       \includegraphics[width=8.8cm,trim={1.5cm 12.9cm 14.3cm 0.5cm},clip]{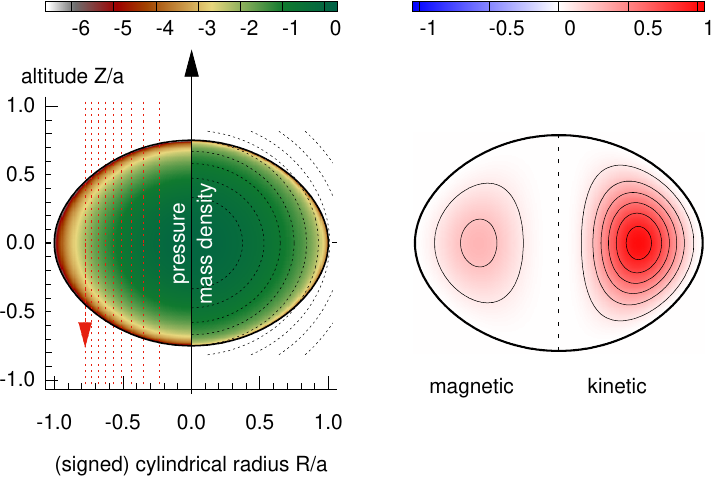}
    \includegraphics[width=8.8cm,trim={1.5cm 13.2cm 14.3cm 1.3cm},clip]{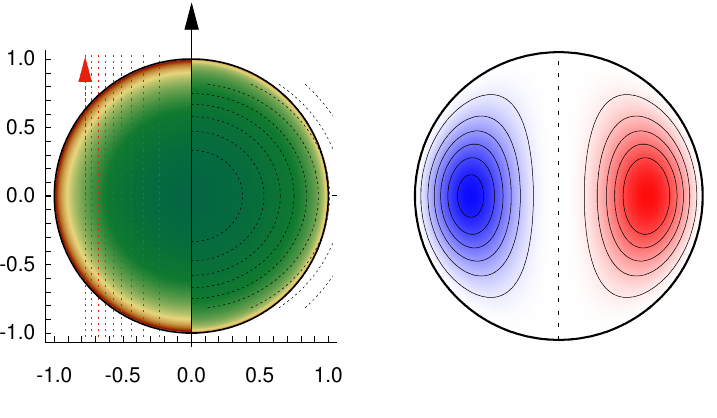}
     \includegraphics[width=8.8cm,trim={1.5cm 13.2cm 14.3cm 1.3cm},clip]{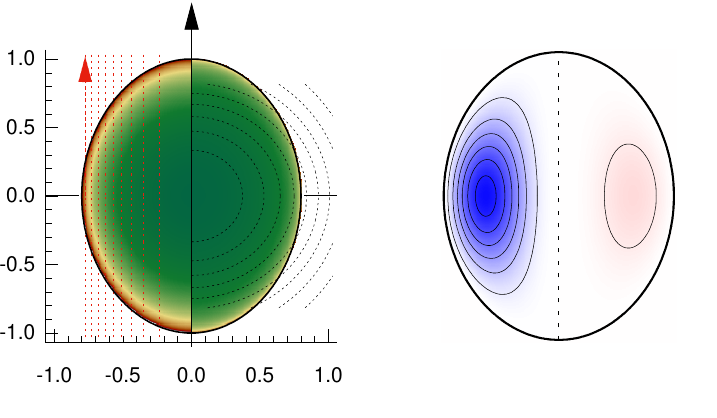}
     \caption{Normalized pressure and mass density in log. scale ({\it left panels}) computed with the {\tt DROP}-code for a polytropic exponent $\gamma=4/3$ (the computational grid has $257 \times 257$ nodes). Simulations differ by the eccentricty and $\eta$-parameter (see also Tab. \ref{tab:data}). The magnetic potential ({\it red lines}) and the gravitational potential ({\it dotted lines}) are superimposed. The arrows give the orientation of $\mathbf{\Omega}$ ({\it black}) and  $\mathbf{B}$ ({\it red}). Also shown is the magnetic energy density relative to the kinetic contribution ({\it right panels}). }
       \label{fig:case2highB_oblate.pdf}
\end{figure}

The expansions above concern the rigidly rotating, homogenous spheroid. The compressibility of matter affects the stratification and shape of the body, which slightly deviates from a spheroid \citep[e.g.][]{hachisu86}. As a matter of fact, $\omeff^2$ has nothing to do with the equation-of-state of matter, which is just incorporated in the first term in Eq.\eqref{eq:bernoulli_bfield}. The statements made above are therefore expected to qualitatively hold for a rigidly rotating, compressible gas as long as {\it the ionisation fraction and the $\delta$-parameter are uniform inside the structure}. We show in Fig. \ref{fig:case2highB_oblate.pdf} three equilibrium configurations computed with the {\tt DROP} code that solves the full $2$D-problem from the Self-Consistent-Field method \citep{hh17}. In all examples, the gas obeys a polytropic equation of state (i.e. the pressure is of the form $p \propto \rho^\gamma)$, and the exponent is $\gamma=\frac{4}{3}$ typical of a fully convective gas. The importance of the magnetic force with respect to the centrifugal force is also shown. The main outputs are listed in Tab. \ref{tab:data}. Selected values for $b/a$ are deliberately very different from unity, in order to make the deformations visible. In the first example (top panels), the axis ratio is $b/a=0.75$ and $\eta <0$. The magnetic field is anti-parallel to the rotation vector $\bm{\Omega}$. The Lorentz force reinforces the centrifugal effect, and rotation is lower than in the non-magnetic case. Self-gravity alone produces the confinement. In the second case (middle panels), we impose $\omeff^2=0$ (i.e. $\eta \approx 1$) and $\varepsilon=0$. With $\delta \bm{B}\cdot\bm{\Omega}>0$, the Lorentz force goes against centrifugation, and the magnetic contribution to the energy density is now negative. The third example is a marked prolate case obtained for $\eta > 1$ and $b/a=1.25$, which corresponds to $\varepsilon^2= - 0.5625$. Here, rotation is weak, and the magnetic field contributes efficiently in the elongation.

    \begin{table}[h]
  \caption{Data associated with the three equilibria displayed in Fig. \ref{fig:case2highB_oblate.pdf}. The solution without magnetic field is in row 2. The kinetic-to-gravitational energy ratio is denoted $T/|W|$, the kinetic-to-magnetic energy ratio is $T/\Gamma_B$, and VP is the Virial parameter.}
  \label{tab:data}
  \centering
    \begin{tabular}{ccccc}
     axis ratio $\bare$ $\blacktriangleright$ &   $0.75$  &       $1$                 &$1.25$\\\hline
  $\bullet$ {Maclaurin ($B=0$)}\\   $\Omega^2/2\pi G\rho$ &    $+0.036068$ &   $-3 \times 10^{-7}$  &    no solution \\   \hline
  $\bullet$ {with magnetic field}\\
   $y \equiv \delta\omb/\sqrt{2 \pi G \rho}$ &    $-0.05$     &   $+0.10$             & $+0.60$ \\
      $\eta$ &   $-0.300202$ &   $+1.000035$         & $+6.607532$ \\
      $\omeff^2/2\pi G \rho$ &   $+0.036068$ &   $-3 \times 10^{-7}$  & $-0.046237$\\
      $\Omega^2/2\pi G \rho$ &   $+0.027740$ &   $+0.999929$         & $+0.008245$\\
     $M/\rhoc \re^3$ &  $+0.430265$ &    $+0.699205$          & $+0.476558$\\
     $T/|W|$ & $+0.027414$ &    $+0.010724$          & $+0.007706$\\
     $T/\Gamma_B$ & $+1.665540$ & $-0.499982$ & $-0.075671$\\
     VP $(\times 10^{-5}$) & $-1.55$ & $-1.26$ &  $-1.13$\\\hline
    \end{tabular}
    \end{table}
    
\section{Summary, validity and discussion}

\begin{figure}[h]
       \centering
       \includegraphics[width=8.7cm,trim={0.4cm 0.cm 0.cm 0.cm},clip]{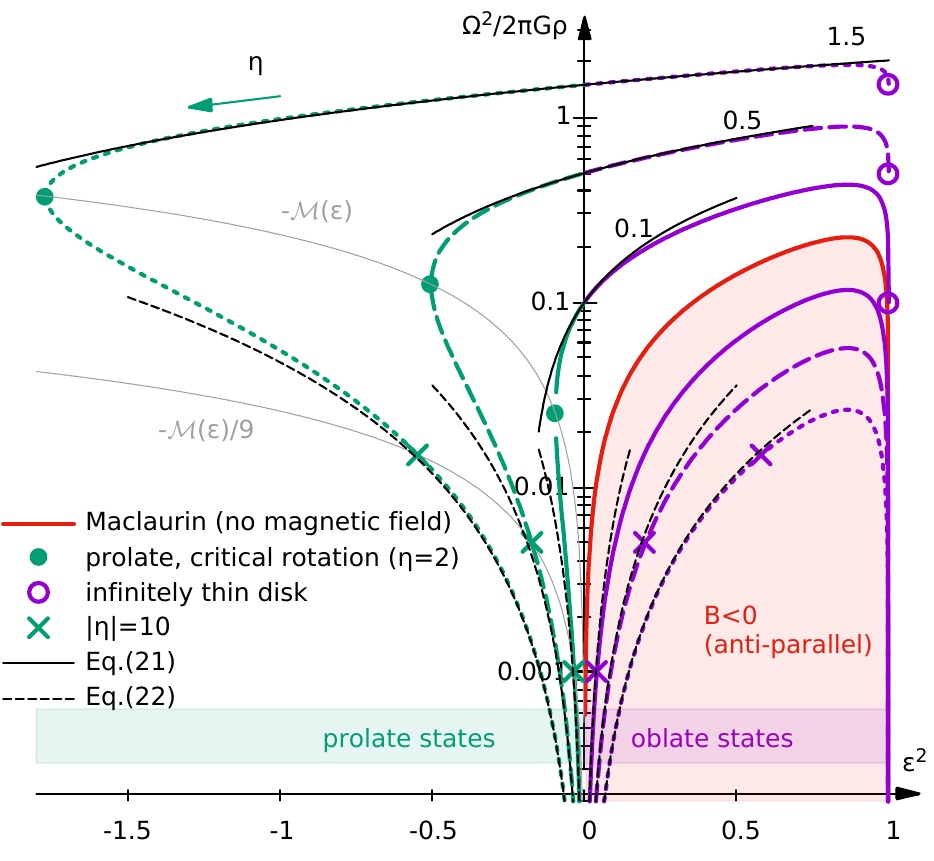}
       \caption{Equilibrium diagram $\Omega^2/2 \pi G \rho$ with magnetic field  in log. scale versus $\varepsilon^2$ for $y^2 \in \{0.1, 0.5,1.5\}$. The Maclaurin function is plotted ({\it red}), as well as the two approximations \eqref{eq:approxhigh} and \eqref{eq:approxlow}. Points where $|\eta|=10$ are indicated ({\it crosses}).}
       \label{fig:newml.pdf}
\end{figure}

Inside a homogeneous spheroid threaded by a coaxial, uniform magnetic field, the total force acting on rigidly rotating azimuthal currents is linear in the cylindrical coordinate, making the configuration a possible figure of equilibrium \citep[e.g.][]{ktye11}. We note that, as any isobar in the body is a spheroid similar to the surface spheroid, {\it this result holds in the presence of a uniform, ambient pressure} $\pamb>0$. In terms of shape, the magnetic states reported in this analysis are indistinguishable from a Maclaurin spheroid. According to Eq.\eqref{eq:om2eff}, the $\Omega^2(\rho,\varepsilon)$-diagram takes a new form 
\begin{subequations}
    \begin{empheq}[left={\empheqlbrace}]{align}
  &\frac{\Omega^2}{2\pi G \rho} = {\cal N}\left(\varepsilon, \frac{\delta \omb}{\sqrt{2\pi G \rho}}\right),\label{eq:newmaclaurin}\\
  &{\cal N}(\varepsilon,y) = \frac{y^2}{4} \left[1\pm \sqrt{1 + \frac{4{\cal M}(\varepsilon)}{y^2}}\right]^2 \label{eq:newmlfunction}
    \end{empheq}
\end{subequations}
where $y^2 \ge -4 {\cal M}(\varepsilon)$. Unsurprisingly, the ionization fraction of matter, the amplitude of the magnetic field and the drift parameter are major ingredients. The function ${\cal N}$ is therefore an extension of Maclaurin's formula. It is displayed versus $\varepsilon^2$ in Fig. \ref{fig:newml.pdf}. We recover ${\cal N}(\varepsilon,0) \equiv {\cal M}(\varepsilon)$ in the absence of any magnetic field or when there is no charge drift (and no net current). Like ${\cal M}$, ${\cal N}$ peaks at $\varepsilon^2 \approx 0.865$, and there are in between $0$ and $3$ states for a given value of $\Omega^2/2\pi G \rho$.  When $\delta\ne 0$, there are two  main families of solutions. For  $|{\cal M}(\varepsilon)/y^2| \ll 1$, we have
  \begin{equation}
    \label{eq:approxhigh}
    {\cal N}(\varepsilon,y) \approx y^2 +2{\cal M}(\varepsilon) \approx y^2+\frac{8}{15}\varepsilon^2 >0,
  \end{equation}
  which approximation is valid whatever the sign of $\varepsilon^2$. This is the upper branch in Fig. \ref{fig:newml.pdf}, where we have a quasi-linear behavior in $\varepsilon^2$ and relatively high rotation rates. The second branch is also twofold as it concerns both oblate and prolate states. It corresponds to
  \begin{equation}
    \label{eq:approxlow}
          {\cal N}(\varepsilon,y) \approx \left[- \frac{{\cal M}(\varepsilon)}{y}\right]^2 \approx \frac{16}{225} \frac{\varepsilon^4}{y^2} >0,
  \end{equation}
which is the lower branch in the figure (low rotation rates). When $\delta \bm{B} \cdot \bm{\Omega}<0$, only oblate shapes are possible. As Fig. \ref{fig:roots.pdf} shows, two cases lead to quasi-spherical structures:
\begin{itemize}
\item $\eta \approx 1$, which means $\Omega \approx \delta \omb$. This case is obtained for various magnitudes of the parameters $B$, $\delta$ and $\xe$, but it requires a specific tunning. This includes states with slow rotation and weak magnetic field, or with fast rotation and strong magnetic field but this requires $\delta\bm{B} \cdot \bm{\Omega}>0$.
\item $|\eta| \gg 1$, which leads to $\Omega \approx - \frac{8 \pi}{15} \varepsilon^2 \frac{ G \rho}{\delta \omb}$. This corresponds to very slow rotation or to large $B$-fields. 
\end{itemize}

As quoted, the novelty is the existence of {\it prolate configurations}, which necessarily occur when the magnetic field has the same orientation as the rotation vector, for a positive drift-parameter. In contrast to oblate configurations (there are solutions for any $\varepsilon \in [0,1]$), prolate states are not always permitted. This is materialized by the existence of critical rotations: i$\varepsilon$, which is a real number, cannot reach infinite values; see Eq.\eqref{eq:crot}. The analog of the flat disk for oblate states, namely the line segment, is not accessible. This phenomenon has already been reported in \cite{ct10}, where the authors have studied an isothermal gas in the context of ideal magneto-hydrodynamics and have accounted for a background magnetic field. We notice, however, that the occurence of critical states does not require any ambient pressure $\pamb$ here, while a link $\bare_{\rm cr}(\pamb)$ is clearly established in \cite{ct10}. It would therefore be interesting to understand the role of compressibility on this phenomenon.

Substantial deformations (oblate or prolate) caused by an external magnetic field are expected as soon as $\omeff$ and $\Omega$ differ significantly from each other, i.e. $|\eta|$ exceeds a few percents typically. By taking $|\eta| > 0.1$ in Eq.\eqref{eq:om2eff}, a natural upper limit for the rotation rate (otherwise centrifugal forces dominates) is 
\begin{flalign}
 \frac{\Omega}{\text{rad/s}} < 9.6 \times 10^4  \times \frac{1}{|\eta|}\frac{|\delta| \xe}{1+\xe}  \left(\frac{|B|}{1 \,{\rm G}}\right)
  \label{eq:omegalupperlimit}
\end{flalign}
where we have set $Z=1$ and $\mio=\mpro$ (proton's mass), and $B$ is in Gauss. As we are dealing with rigid rotations, another constraint is basically imposed by the size of the system, namely $\Omega a < c$, where $c$ is the speed of light.

\begin{figure}[h]
  \hspace*{-0.5cm}\includegraphics[trim={6.5cm 13cm 7.8cm .25cm},clip,width=1.08\linewidth]{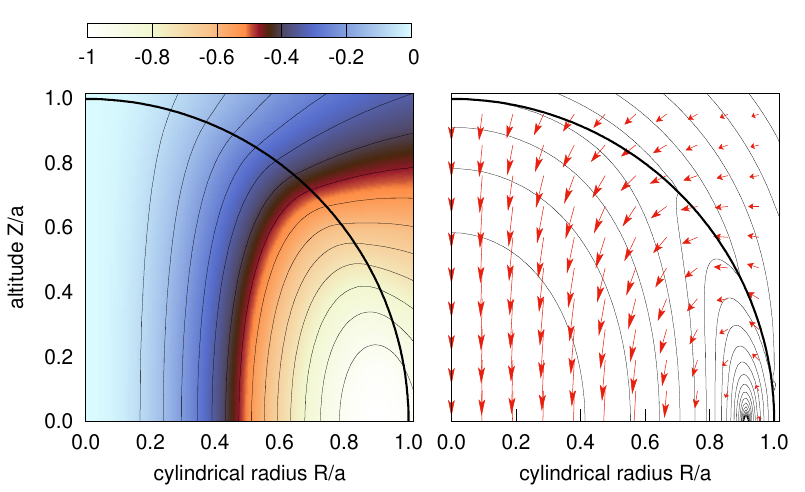}
  \caption{Normalized magnetic potential $\Lambda_{\rm self}$ in color code ({\it left}) and ${\bm B}_{\rm self}$-field ({\it right}) in units of $\delta \Omega$, for a homogeneous spheroid with $\varepsilon=0$ (non-rotating limit).}
\label{fig:ABfield.pdf}
\end{figure}

The validity of the present analysis also rests upon the fact the $B$ is unperturbed by the flow and must overpass the magnetic field ${\bm B}_{\rm self.}$ produced by the rotating charges, which must be checked even if the ionisation fraction is low. We have therefore calculated the potential vector ${\bm A}_{\rm self.}$ associated with the azimuthal current ${\bm j}$ defined by Eqs.\eqref{eq:jcurrent} and \eqref{eq:assumptiondrift}, from the technique reported in \cite{m82}. The magnetic potential $\Lambda_{\rm self.}=R {\bm A}_{\rm self.} \cdot {\bm e}_\phi$ and the corresponding poloidal field ${\bm B}_{\rm self.} = \nabla \times {\bm A}_{\rm self.}$ are shown in Fig. \ref{fig:ABfield.pdf} in units of $\delta \Omega$ in the spherical limit. It is worth noting that, in contrast with the assumption retained in \citep{ktye11}, the induced field is far from purely vertical everywhere in the system. At the center, the field is relatively large and estimated to $-\frac{1}{3} \mu_0 \nel e \delta \Omega a^2$. Taking this value as the typical magnitude for $B_{\rm self.}$, the condition of self-consistency $|\bself| \ll |B|$ reads
\begin{equation}
\frac{1}{\eta} \ll \frac{8}{\pi^2} \left(\frac{4 \pi \varepsilon_0 G \mio ^2}{e^2}\right) \left(\frac{Z+\xe}{\delta Z\xe}\right)^2 \left(\frac{c}{a \Omega_{\rm ff}}\right)^2
     \label{ineq:limit}
\end{equation}
where $\sqrt{G \rho}$ has been written in terms of the free-fall time $1/\Omega_{\rm ff}$. Still for a hydrogen gas ($Z=1$ and $\mio = \mpro$),
 this inequality reads
\begin{equation}
  \label{ineq:eta}
\eta \gg 1.5 \times 10^{36} \left(\beta \frac{\xe}{1+\xe} \delta \right)^2 \equiv \eta_0,
\end{equation}
where $\beta=\Omega_{\rm ff} a/c \ll 1$ is the relativity factor at the equator. We can see from Fig. \ref{fig:newml.pdf} that highly elongated/prolate shapes are accessible for very large $\eta$'s, which requires for instance very low rotation rates. We notice that the threshold $\eta_0$ can be considerably decreased if simultaneously $\beta^2 \ll 1$, $\xe^2 \ll 1$ and $\delta^2 \ll 1$. It follows that the solutions discussed here make sense preferentially in {\it non-relativistic and weakly ionized rotating systems, and for small electron/ion relative drifts.}

Which systems are potentially concerned ? The structure of an ``isolated'' star cannot be regulated, even partially, by an external magnetic field. The impact of magnetic fields on the structure of normal and compact stars is studied for long, and widely documented \citep[e.g.][]{s49,m65,cm71,m75,horedttextbook2004,lj09,ugmfte14,p22}. These fields are linked to the toroidal and poloidal circulations of the stellar plasma, which is a complicated problem \cite[e.g.][]{cow53,mr62}. Galactic field strengths are typically $6$ to $9$ orders of magnitudes below what conducting plasmas in stars can generate \citep{bbmss96}. White dwarfs and neutron stars have huge surface magnetic fields, from $10^7$ to $10^{15}$ G typically and therefore represents very powerful dipoles capable of influencing their environment. In double systems, the structure and spin of the secondary component (a giant star or a even planet), in particular the upper-layers, can probably be impacted, depending on the orbital separation and conductivity of matter \citep{c01,ibsh14,pk17}. This would merit further investigations.

    \begin{figure}
  \includegraphics[trim={6.5cm 13cm 14.1cm 0.cm},clip,width=5.2cm]{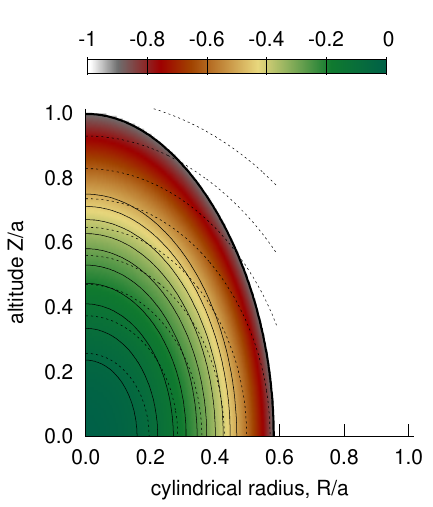}
  \caption{Mass-density profile  $\rho(R,Z)$ for a typical core obtained with the {\tt DROP}-code for input parmeters listed in Tab. \ref{tab:acore}. The gravitational potential is also shown ({\it dotted lines}).} 
\label{fig:acore.pdf}
\end{figure}

    \begin{table}
  \caption{Data associated with the configuration shown in Fig. \ref{fig:acore.pdf}; see also \cite{ct10}. Values of $\bself$ correspond to the vertical component at the center. The ionisation fraction is calculated according to \citep{hpfl22}. The energy densities are in units of $G \rhoc^2 a^5$.}
  \label{tab:data}
    \begin{tabular}{rlc}\\
 {\bf input}&  equatorial radius $a$ & $0.14$ pc \\
     & axis ratio $b/a$ & $\frac{1}{0.582} \approx 1.718$ \\
     & \qquad \rotatebox[origin=c]{180}{$\Lsh$} eccentricity $\varepsilon^2$ & $-1.952$ \\
     & ambiant pressure $p_{\rm amb}$ & $10^{-12}$ dyn/cm$^2$\\
     & surface mass density $n_{\rm amb}$ & $1500 \, \mpro$ /cm$^3$\\
     & mass density ratio $\rho_{\rm amb}/\rhoc$ & $\frac{1}{2.5}$\\
     & magnetic field $B$ & $20$ $\mu$G \\
 {\bf additional} & polytropic exponent $\gamma$ & $4/3$ \\
 {\bf param.}    & ionisation fraction $\xe(\rhoc)$ & $5.474 \times 10^{-8}$ \\
     & $y = \delta \omb/\sqrt{2\pi G \rhoc}$ &   $1.079$ \\\hline
 {\bf output} & $\ombo$ &  $1.912$ rad/s \\
     & $\omb$ & $1.047 \times 10^{-7}$ rad/s\\
     & drift parameter $\delta$ & $10^{-8}$ \\
     & $|\bself|$ & $1.457$ $\mu$G \\
 & rotation rate $\Omega$      & $3.158 \times 10^{-14}$ rad/s\\
 & $\sqrt{\partial P /\partial \rho}$ & $0.264$ km/s\\
     & $\beta=\Omega a/c$ & $ 4.541 \times 10^{-7}$\\
     & $\eta$ &   $1.753$  \\
     & $\omeff^2/2\pi G \rhoc$ &   $-0.285$ \\
     & $\Omega^2/2\pi G \rhoc$ &   $+0.379$ \\
     & total mass $M$ &  $1.089$ M$_\odot$ \\
 & kinetic energy $T$ & $1.825$\\
 & gravitational energy $W$ & $-9.627$\\
 & internal energy & $+24.093$\\
 & magnetic energy & $-6.398$ \\
 & ambient pressure contrib. & $-11.711$\\
 & $T/|W|$ & $+0.189$ \\\hline
      \multicolumn{3}{l}{{\bf magnetic Maclaurin} (with $\rho=\rhoc$)}\\
      & $\eta_0$, from Eq.(23) & $2.593 \times 10^{-4}$\\
     & $|\bself|$ & $1.833$ $\mu$G\\\hline
    \end{tabular}
    \label{tab:acore}
    \end{table}
    
In dense interstellar clouds and prestellar cores, the magnetic field plays a critical role and regulates star formation \citep{c12}. Line-of-sight estimates give $10 \, \mu$G typically \citep{gchmt89}, and the kinematics in some clouds seem compatible with rigid rotation \citep{gbfm93}. The ionization fraction is known to be very low \citep{cwth98,cwztm02}. Values for $\xe$ depend on the mass-density, chemical content, dust grains and ionisation rate by cosmic rays \citep[e.g.][]{od74,wfd04,bron21}, and are typically in the range $10^{-4} - 10^{-9}$. With the prescription for $\xe(\rho)$ established in \cite{hpfl22}, Ineq.\eqref{eq:omegalupperlimit} reads\footnote{ This formula, based on Eq. \eqref{eq:omegalupperlimit}, takes into account molecular hydrogen H$_2$ as the dominant species and helium in cosmic proportions, and we have $\rho \approx 2.33 \, \nel \mio/\xe$ and $\no=0.6 \, \npro$.}
\begin{flalign}
  \frac{\Omega}{\text{rad/s}} \lesssim 2.7 \times 10^{-9}  |\delta| \left(\frac{\npro}{10^4 \, {\rm cm}^{-3}}\right)^{-\frac{1}{2}} \left(\frac{|B|}{1 \,\mu {\rm G}}\right)
  \label{eq:omegalupperlimit2}
\end{flalign}
where $\npro$ is the total number density of H-atoms. The normalizations adopted here seem to be lower limits, both in terms of number density and magnetic fields, and can be both $1000$ times larger typically \citep{hen18,hz19,pftln23}. Ineq.\eqref{eq:omegalupperlimit2} seems fully pertinent to prestellar cores, which are believed to have a low rotational support. The key-point is the $\delta$-parameter, which should be very small in the interstellar medium. With the same scalings as above and for a typical size $a=0.1$ pc, we get from Maxwell Ampere's law $V_e-V_i \sim \frac{B}{\mu_0ea \npro} \approx 0.16$ cm/s, while the isothermal speed at $10$ K is the order of $3 \times 10^4$ cm/s. If this value is confunded with the bulk velocity (the sonic limit), then we find $\delta \approx 5 \times 10^{-6}$, but we are confident that this estimate is poorly reliable.\\

 We show in Fig. \ref{fig:acore.pdf} the mass-density profile computed with the {\tt DROP}-code for a typical interstellar core scaled to about $0.1$ pc. This example corresponds to a heterogenous gas, assuming a ``soft'' polytropic equation of state. It includes ambient pressure, and we take $\delta=10^{-8}$. The input and output data are listed in Tab. \ref{tab:acore}. This state is close to a critical prolate equilibrium. This simulation compares very well with the results obtained in \cite{ct10} with similar parameters. The authors have observed a limit in the diameter of structures, which property is also found here, not only in the homogeneous limit but also in the case of a compressible gas.\\

This study opens onto interesting problems. This is for instance the determination of equilibrium sequences in compressible cases, by varying for instance the eccentricity $\varepsilon$ for fixed values of $\eta$ and $\omb$, and the systematic determination of the critical prolate states \cite{ct10}. Equations and limits derived in this article are actually specific to the homogeneous case. Due to the magnetic support, we expect that the endings of compressible oblate sequences of equilibrium are changed \citep{hachisu86} in terms of $\Omega$, again depending on $B$. It would be necessary to analyze the stability of these configurations, possibility leading to constraints on the external magnetic field, ionsation fraction and drift parameter. Another interesting point concerns the determination of configurations if $\bm{B}$ and $\bm{\Omega}$ are misaligned \citep[e.g.][]{y21,kt21,pftln23}, which requires a tridimensional, numerical treatment. Finally, a key-question concerns the link between the kinematics of charged particles and neutrals, which is certainly more complex than considered here, and which might invalidate the present approach. It is clear that the drift-parameter, which is plausibly much lower than the sound speed by orders of magnitude, is hard to guess and remains a ``big'' unknown here, as quoted above.

\begin{acknowledgments}
  We are grateful to our colleagues T. Csengeri, B. Commer\c{c}on, B. Godard, E. Gourgoulhon, S. Lander, F. Le Petit, G. Pineau-des-For\^{e}ts for fruitful inputs and references, and especially J. Ferreira for key-advices about the single-fluid approach. We thank the referee for interesting and constructive criticisms.
\end{acknowledgments}

\bibliographystyle{apsrev4-1}
%


\end{document}